\documentclass[reprint, amsmath,amssymb, aps]{revtex4-1}
\usepackage{graphicx}
\usepackage{dcolumn}
\usepackage{bm}
\begin{document}

\preprint{APS/123-QED}

\title{Varistor characteristics of a nano-device containing graphene and oxidized graphene: Verification by DFT + NEGF}

\author{Badie Ghavami}
\email{badie.ghavami@azaruniv.edu}
\author{Alireza Rastkar-Ebrahimzadeh}
 \email{a\_rastkar@azaruniv.edu}
\affiliation{Molecular Simulation Laboratory, Department of Physics, Faculty of Basic Sciences, Azarbaijan Shahid Madani University, Tabriz, Iran. Fax: +98-41-34327541; Tel: +98-41-34327541}

\date{\today}

\begin{abstract}
Electron transport and quantum conductance through an armchair graphene and its oxidized graphene- containing form were investigated by the density functional theory (DFT) method and the implementation of the non-equilibrium Green function (NEGF) approach. The computed $I-V_b$(current as a function of bias voltage) characteristic of the studied systems showed the tunneling phenomenon in bias and gate voltages considered. Along with the transport properties, electronic properties including density of states (DOS) were calculated in the studied systems. A close examination of the results showed that the $I-V_b$ curve for graphene behaved $I\propto V_be^{\lambda V_b}$ like  at some bias voltages, while for the oxidized graphene-containing form, its trend was the same as that of a Voltage Dependent Resistor (VDR-VARiable resISTOR), $I\propto V_b^\beta$, at the whole range  of the applied bias.
\begin{description}
\item[Usage]
Secondary publications and information retrieval purposes.
\item[PACS numbers]
May be entered using the \verb+\pacs{#1}+ command.
\item[Structure]
You may use the \texttt{description} environment to structure your abstract;
use the optional argument of the \verb+\item+ command to give the category of each item. 
\end{description}
\end{abstract}

\pacs{Valid PACS appear here}
\maketitle


\section{\label{sec:level1}Introduction}
\begin{figure}
\includegraphics[height=4cm]{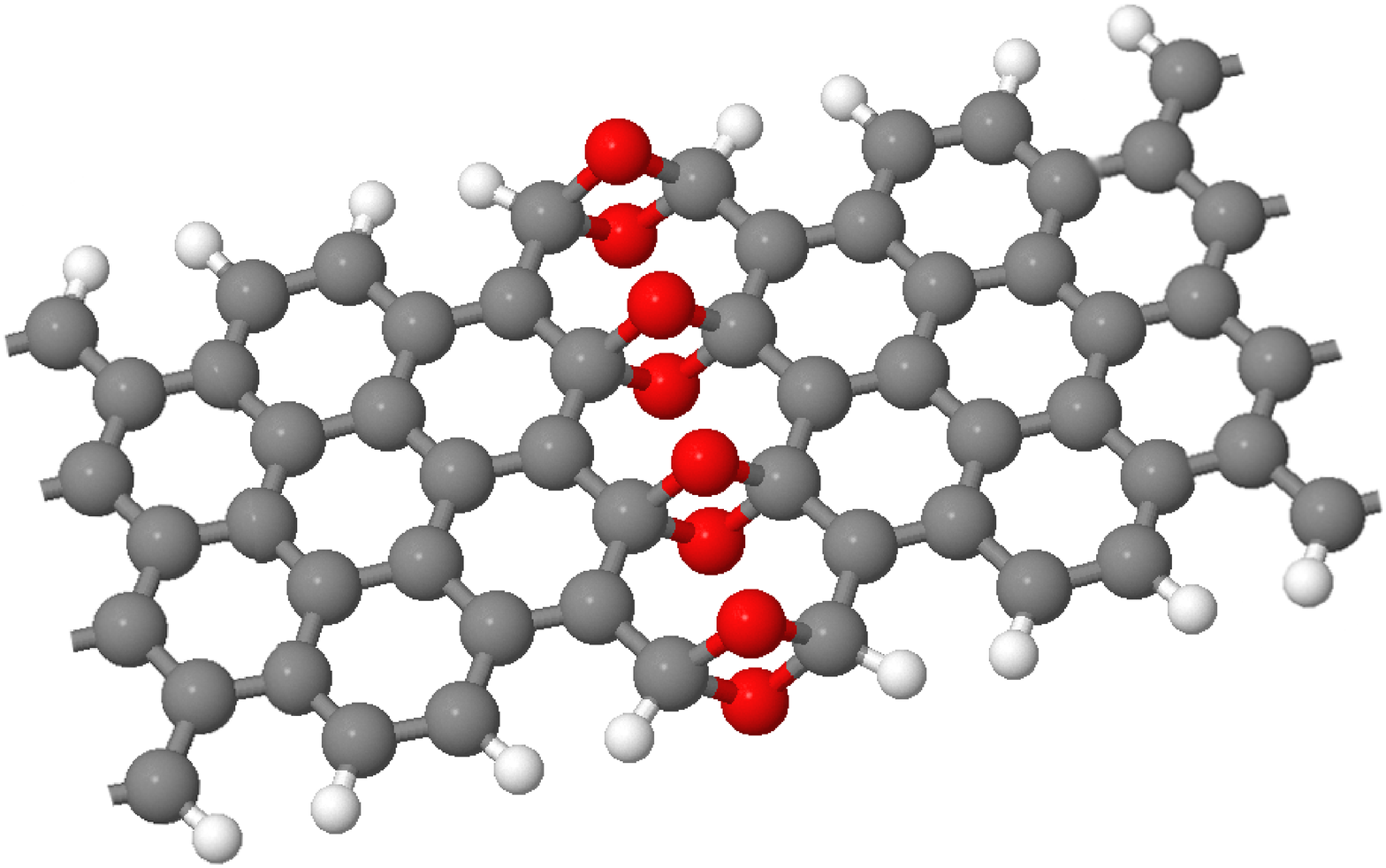}
\includegraphics[angle=0, width=0.4\textwidth]{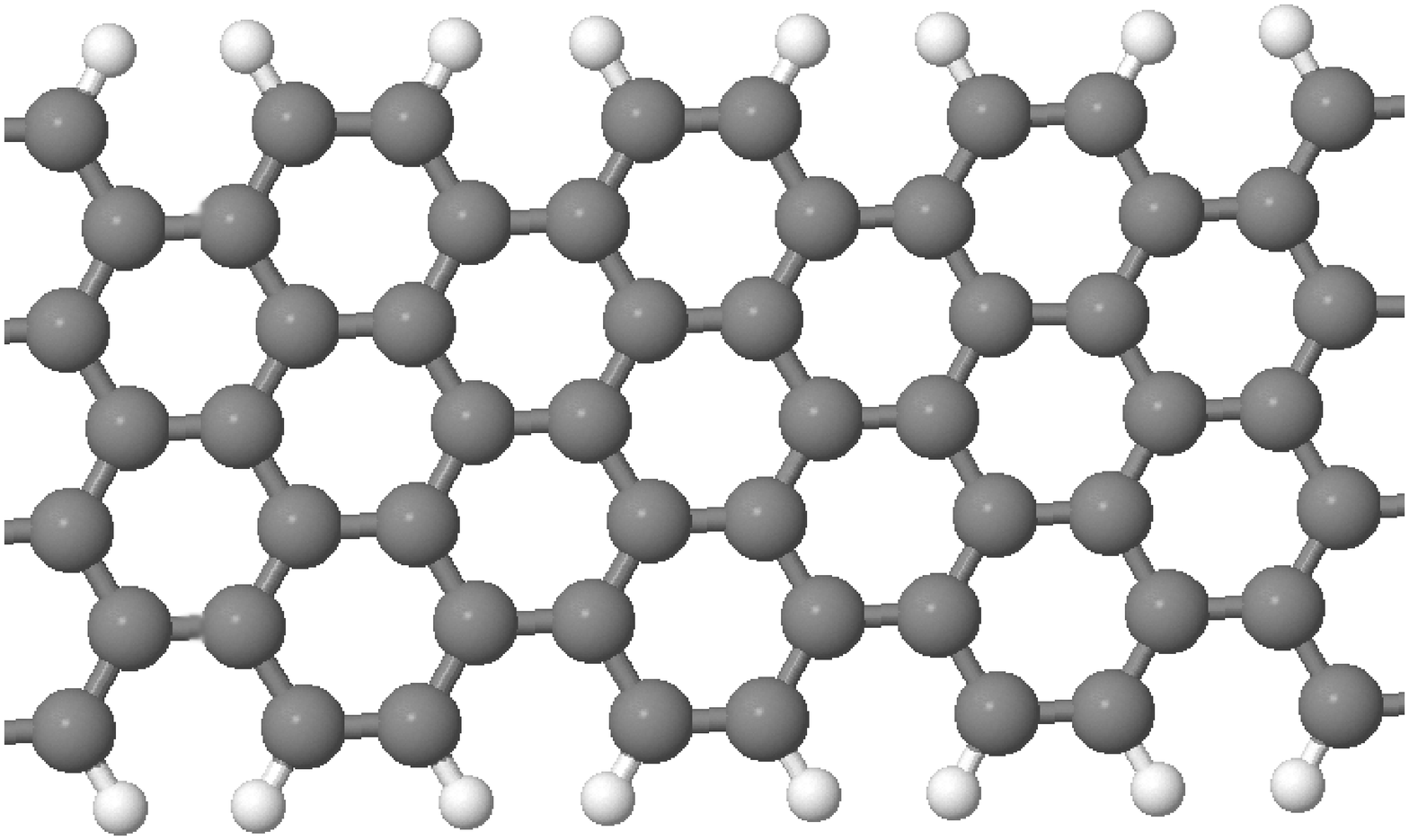}
\caption{\label{fig:1} Considered (a) oxidized graphene- containing and  (b) graphene systems.}
\end{figure}

The equilibrium electronic properties and electronic transport properties are important in designing and manufacturing novel nano-electronics devices. Recently, several interesting review articles have been published\cite{Zhong,Savage,David,Zhu,Lherbier,Prasongkit,Gruznev,Benchamekh,Zhen} in this challenging field, covering theoretical physics to organic chemistry. Graphene is a perfect $sp^2$-hybridized carbon mono-layer sheet that has attracted much attention since its discovery\cite{Geim,Novoselov,Motta}. Also, as a theoretical interest, its possible applications in carbon-based electronic structures and conductivity can be considered. Graphene has a very peculiar electronic structure, as it is geometrically symmetrical and the electrons in the two dimensions are confined. Graphene is a semi-metal whose specific linear electronic band dispersion near Dirac points (the Brillouin zone corners) give rise to holes and electrons propagated as massless fermions\cite {Lherbier1,Charlier, Castro, Wallace, Luis}. Graphene nanoribbons (GNR) are structural derivatives of grapheme that are considered a promising candidate as the building devices for future electronic applications\cite{Wakabayashi,Avouris,Wang,Sung}. The unusual semi-metallic behavior of graphene has been discovered in the computational work of Wallace\cite{Wallace}. Quantum interference phenomena such as universal conductance fluctuation and weak localization or Aharonov-Bohm effect in graphene rings have been shown experimentally\cite{Morozov, Recher, Russo}. The properties of graphene could be varied by structural variations applied on it, such as doping\cite{Yan,Riedl,Kaukonen} or functionalizing\cite{Paul,Yang,Wood}. For instance, functionalization of graphene with oxygen and hydrogen was performed by Lopez-Bezanilla et.al\cite{Bezanilla}. These structural variations were found to affect the equilibrium and transport properties of graphene\cite{Yamacli}. The capability of engineering the electronic transport properties, such as the ballistic electronic propagation and quantum conductance, is useful in manufacturing the field effect devices\cite{Novoselov}. The nonlinear behavior of the current-voltage($I-V_b$) curve or the varistor effect in inhomogeneous materials has been extensively studied for using varistors in limiting the transient over voltage generated by electromagnetic effects\cite{Beneden}. The preparation of an Ag-graphene epoxy nanocomposite with a varistor effect near the percolation threshold was described by Q. Liu et al \cite{Liu} and Lin et al \cite{Lin}. The varistor effect was caused by the intrinsic electrical nonlinearity in the defective graphene sheets with an experimental method. It seems that reversible electrical nonlinearity existing in polymer composites consisting of some conductive filler and an insulating polymer holds promise for use in varistors\cite{Bidadi,Lin}.\\
In this paper, devices composed of armchair pristine graphene nanoribbon or oxidized graphene nanoribbon as the scattering central region and armchair pristine graphene nanoribbon as source and drain electrodes were investigated. The computations were performed in bias voltages ranging from -2.0 to 2.0 eV divided by 0.01 eV intervals and at gate voltages including -3.0, 0.0 and +3.0 eV.
Figures 1-a and 1-b(Fig.~\ref{fig:1}) present the devices formed from oxidized graphene-containing and pristine graphene as the central scattering region, respectively. At the first glance, the considered systems were optimized using the density functional theory (DFT) computations. Then, the transmission coefficient and the electrical current in each bias voltage in the considered gate voltages were computed by utilizing the non-equilibrium green function (NEGF) method. The computational results were analyzed and interpreted with the help of the transmission spectrum and the total and projected density of states for the considered devices. The current-voltage values obtained for the considered devices corresponded to the well-known varistor characteristics. The computed $I-V_b$ behavior for pristine and oxidized graphene- containing devices were compared and discussed in terms of electronic structure variations caused by oxidation of graphene.
\section{\label{sec:level2}COMPUTATIONAL METHOD}
The transmission function of the system was obtained by the following equation\cite{Datta}:
\begin{equation}
T(E,V)=Tr\left[\Gamma_L(E,V)G(E,V)\Gamma_R(E,V)G^\dagger(E,V)\right] 
\end{equation}
where $E$ and $V$ are energy and bias voltage, respectively. Also $\Gamma_{R(L)}$ is the spectral density describing the coupling between the right(left) electrode and the scattering region. In this equation, $G(E,V)$ is the green function formally given by:
\begin{equation}
G(E,V)=\frac{1}{\left[ES-H(V)-\Sigma_L(E,V)-\Sigma_R(E,V)\right]}
\end{equation}
where $S$ is the overlap matrix, $H(V)$ and $\Sigma_{R(L)}(E,V)$  are the Hamiltonian of the system and the self-energy of lead right(left) when a bias voltage is applied. Spectral density is given by the imaginary part of the electrode self-energy($\Sigma_{R(L)}$):
\begin{equation}
\Gamma_{R(L)}(E,V)=i\left(\Sigma_{R(L)}-\Sigma^{\ast}_{R(L)}\right) 
\end{equation}
The implementation of gate voltage on the systems was treated by adding
an electric potential defined by\cite{Ozaki5}:
\begin{equation}
V_g(x)=V_g^{(0)}exp\left[-\left(\frac{x-x_c}{d}\right)^8\right] 
\end{equation}
where $V_g$, $x_c$ and $d$  are a constant value corresponding to the gate voltage,  the center of the scattering region, and the length of the unit vector along the X axis for the scattering region. The electric potential may resemble the potential produced by the image charges\cite{Liang}.
The current was given by the Landauer Buttiker\cite{Landauer, Buttiker} formula, which is the following integral:
\begin{equation}
I(V)=\frac{2e}{h}\int_{-\infty}^{+\infty}dET(E,V)\left[f(E-\mu_L)-f(E-\mu_R)\right]
\end{equation}
where $\mu_{R(L)}$ is the chemical potential of the right(left) electrode, which is $eV=\mu_L-\mu_R$. Also, $f(E-\mu_{R(L)})$ is the occupation Fermi function.
All calculations were performed by the OPEN source Package for Material eXplorer version-3.7 (OPENMX-3.7) computer package code\cite{OPENMX}. This package uses PAOs centered on atomic sites as the basis functions\cite{Ozaki3, Ozaki6} generated by a confinement  scheme\cite{Ozaki2,Ozaki6}. In all of the DFT computations, local density approximation, LDA, was employed as the exchange-correlation functional\cite{Ceperley, Perdew}.\\
\section{\label{sec:level3}RESULTS}
\begin{figure}
\includegraphics[height=7cm]{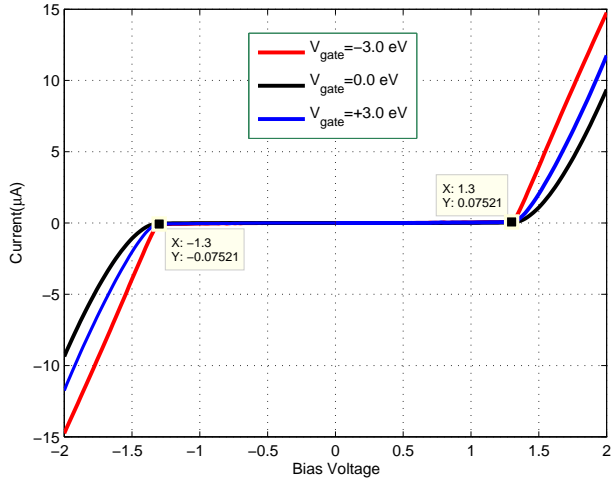}
\caption{\label{fig:2} Current versus bias voltage at considered gate voltages for graphene.}
\end{figure}

\begin{figure}
\includegraphics[height=7cm]{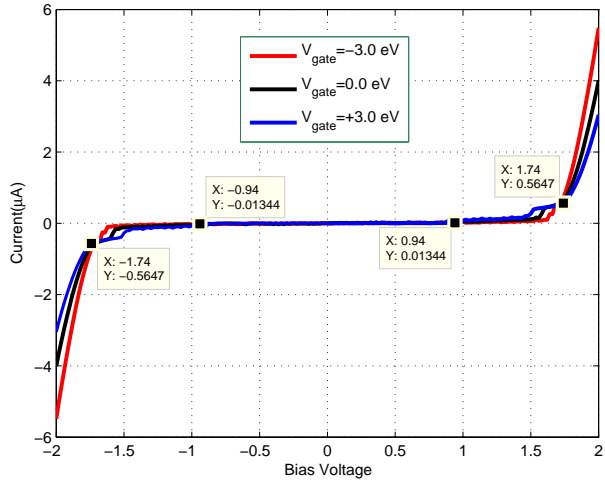}
\caption{\label{fig:3}Current versus bias voltage at considered gate voltages for oxidized graphene-containing device.}
\end{figure}

The current-bias voltage values for the studied pristine and oxidized graphene containing devices have been presented in Fig.~\ref{fig:2} and Fig.~\ref{fig:3}, respectively. It is clear from these figures that there was a considerable similarity between the $I-V_b$ characteristics of the studied systems. It can also be seen that in biases lower than 1.3 eV and 0.94 eV in the graphene and oxidized graphene-containing, the respectability of the current was essentially zero. However, at the greater bias voltages, the current was increased with a relatively sharp slope for the applied gate voltages. Along with the similarities in $I-V_b$ characteristics of the studied structures, there was a major difference in currents value: the current of pristine graphene was twice more than that of the oxidized graphene current. This made the oxidized graphene a complete semiconductor against the semi-metal pristine graphene. The advantages of the oxidized graphene, compared to other semiconductors, could be understood from these figures, which show varistor\cite{resistor} properties. The current of the oxidized graphene device was lower than that of graphene device in the considered bias range. This was probably due to different hybridizations schema in the considered structures. The $sp^2$ hybridization in graphene was changed to $sp^3$ in the oxidized graphene and consequently, the $\pi$ delocalized molecular orbitals were destroyed. Therefore, a main contribution in electronic transport was vanished in the oxidized graphene.
The computed currents at the considered gate voltages have been presented in Fig.~\ref{fig:2}. It can be seen that the current at the zero gate voltage was the lowest compared to gate +3 and -3 eV gate voltages. This figure also shows that at the +3 eV gate voltage, the current had lower values, in comparison to -3 eV gate, for the bias voltages 1.3 to 2.0 eV.  Figure 4 also shows the density of states for the graphene device along with the transmission spectrum. This figure shows that at the zero gate voltage, graphene had the lowest density of states, causing the decrease of current. Furthermore, the displacements Fig.~\ref{fig:2} occurring in the energy levels of the scattering region were raised from the applied gate voltage\cite{Datta}, thereby leading to increasing or decreasing the current. The current rose when the energy levels of the scattering region was between $\mu_L<\epsilon<\mu_R$\cite{Datta}. Therefore, the variations in the current values observed in Fig.~\ref{fig:2} could be attributed to the displacement of energy levels of the scattering region due to changing the gate voltage from +3 to -3 eV.
Fig.~\ref{fig:3} shows current versus bias voltage at the considered gate voltages for the oxidized graphene- containing device. This figure shows three regions: i) the first region is one in which the current is zero (corresponding to the bias values of 0.0 to 0.94 eV), ii), the second is one in which the current increases slightly (corresponding to the bias values from 0.94 to 1.74 eV) and iii) the third region is one in which the current increases sharply (corresponding to the bias values from 1.74 up to 2.0 eV).  This figure also shows that the bias voltage equal to 1.74 eV is a critical point in which the currents belonging to the gate voltage of -3 eV and +3 eV are displaced beyond it. The DOS and transmission spectrum of oxidized graphene- containing device can also be observed. The figure also indicates that the DOS of oxidized graphene- containing in had lower values in the gate voltage of -3 eV, compared to zero gate voltage, and it was also lower than +3 eV gate voltage. In other words, the current in the zero gate voltage lied between the currents in the gate voltage of +3.0 and -3.0 eV. A close examination of the results showed that we could correlate the computed $I-V_b$ data using well-known exponential and  power-low equations in terms of the bias voltage, at bias voltages greater than 1.3 eV for oxidized graphene-containing and in all bias voltage range for pristine graphene. The equations used in order to correlate the $I-V_b$ values have the following forms:
\begin{equation}
I=\gamma V_be^{\lambda V_b}
\end{equation}
\begin{equation}
I=\alpha V_b^\beta
\end{equation}
where $\alpha$  and $\beta$  are adjustable parameters for oxidized graphene- contacting and $\gamma$ and $\lambda$ are pre-exponential and exponential adjustable parameters for graphene. Recently, A. B. Kaiser et al.\cite{CP723} have empirically verified that these equations represent experimental current-bias voltage values measured for the network of Ag−V2O5 nanofibres at various temperatures. Also Q. Liu et al\cite{Liu} and  Z. Brankovic et al \cite{ZnO} have correlated the experimental $I-V_b$  data of Ag-graphene epoxy composite and ZnO devices using similar power-low equations in terms of bias voltage.\\
\begin{table}
\caption{\label{tab:table1}
Adjustable fitted parameters of equation\\
 $I=\gamma V_bexp(\lambda V_b)$ for graphene.}
\begin{ruledtabular}
\begin{tabular}{lcdr}
\textrm{System}&
\textrm{$\gamma$}&
\textrm{$\lambda$}&
\textrm{RMS}\\
\colrule
Gate(-3.0 eV) & 21.7295 & 0.0000  & 0.0092  \\
Gate(0.0 eV) & 8.9841 & 0.9119 & 0.1027  \\
Gate(+3.0 eV)  & 14.1723 & 0.4593 & 0.0315  \\
\end{tabular}
\end{ruledtabular}
\end{table}
\begin{table}
\caption{\label{tab:table2}
Adjustable fitted parameters of equation\\
 $I=\alpha V_b^\beta$ for oxidized graphene-containing device.}
\begin{ruledtabular}
\begin{tabular}{lcdr}
\textrm{System}&
\textrm{$\alpha$}&
\textrm{$\beta$}&
\textrm{RMS}\\
\colrule
Gate(-3.0 eV) & 0.0005& 13.2600 &  0.0952  \\
Gate(0.0 eV) & 0.0009 & 12.1600& 0.0518  \\
Gate(+3.0 eV) & 0.0023 & 10.3200 & 0.0776 \\
\end{tabular}
\end{ruledtabular}
\end{table}
Table~\ref{tab:table1} and Table~\ref{tab:table2} include the values of adjustable parameters obtained by a non-linear least squares regression method along with the standard deviation in the current. It is also shown that the beta parameter in the case of graphene, at -3 eV gate potential, gets a zero value. This means that this device has ohmic behavior and the current varies linearly with the bias voltage at this range. The results for the oxidized graphene-containing showed that the studied systems could be used as a Voltage Dependent Resistor (VDR-VARiable resISTOR: VARISTOR)\cite{resistor} in nano- electronics devices.\\
\begin{figure}
\includegraphics[height=7cm]{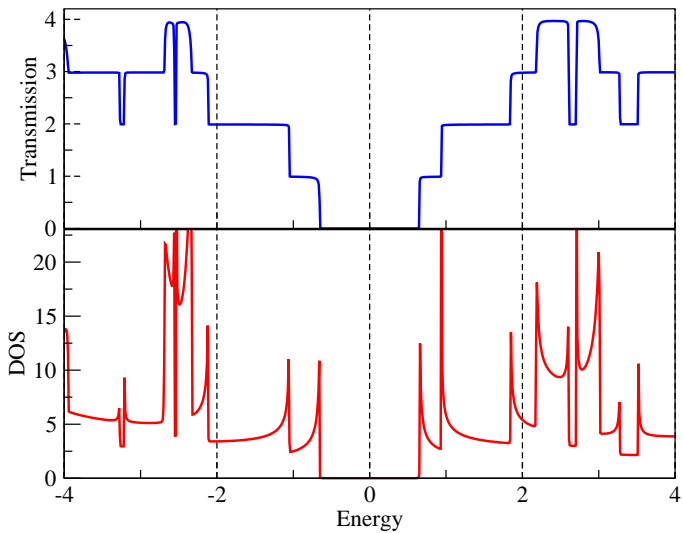}
\caption{\label{fig:4} Transmission spectrum and DOS at zero bias and gate voltages for graphene.}
\end{figure}
\begin{figure}
\includegraphics[height=7cm]{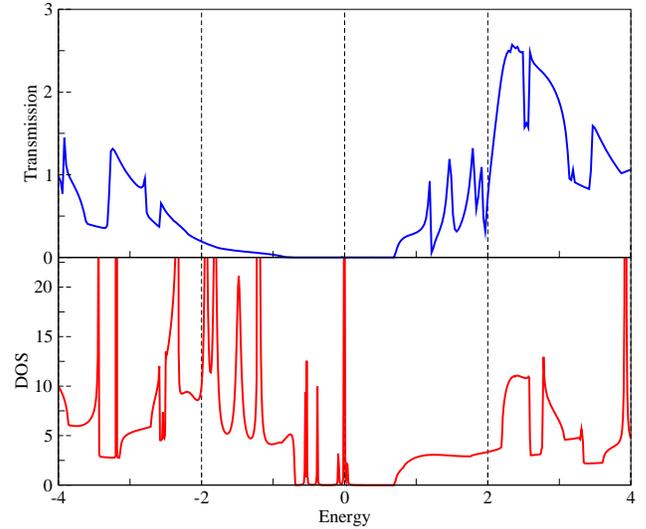}
\caption{\label{fig:5} Transmission spectrum and DOS at zero bias and gate voltages for oxidized graphene-containing device.}
\end{figure}
Transmission spectra along with the DOS of the pristine graphene- containing device have been presented in Fig.~\ref{fig:4}. This figure shows an excellent agreement between transmission and DOS spectroms. Fig.~\ref{fig:5} shows the transmission spectrum and DOS for oxidized graphene-containing device. However, for this system, there were some values of energy (-0.7 to +0.8 eV) for which DOS got non-zero values, but transmission was vanished. This inconsistency between transmission and DOS values could be interpreted with the help of the projected density of states, PDOS, values for device and leads regions. Fig.~\ref{fig:6} shows PDOS of oxidized graphene-containing device and the corresponding leads. This figure also shows that the PDOS of scattering region had some nonzero values in a range wherein PDOS of the leads were vanished. In other words, at the aforementioned energy range (-0.7 to +0.8 eV), the scattering region had some states, but the leads did not. This may be interpreted as an absence of orbital overlap, eigenchannel formation or electron tunneling along the scattering region and leads. These result in a zero transmission for oxidized graphene-containing device in the energies -0.7 up to +0.8 eV.
\begin{figure}
\includegraphics[height=9cm]{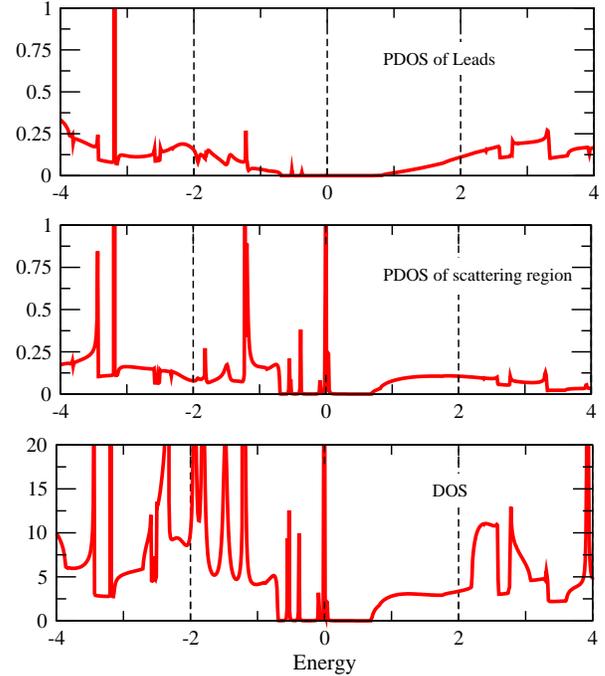}
\caption{\label{fig:6} PDOS of the lead and scattering region for oxidized graphene-containing in zero bias and gate voltages.}
\end{figure}
\section{\label{sec:level4}CONCLUSION}
The electronic structure and electronic transport properties of some devices composed from pristine graphene and oxidized graphene-containing as the scattering region were investigated using the DFT+NEGF method. The computational results including DOS, $I-V_b$ and transmission spectrum were presented, compared and discussed for the studied systems. The computations were done in bias voltages ranging from -2 eV to 2 eV divided by 0.01 eV steps, in the gate voltages of -3, 0.0 and +3.0 eV. The results for current-bias values were compared for the studied systems and correlated using well known empirical equations. These correlations for oxidized graphene-containing form were of power-low form type and described VARISTOR characteristics. Also, the consistency between transmission spectrum and PDOS for the studied systems were analyzed and discussed.
\section{ACKNOWLEDGMENTS}
This research was supported by a research fund "$No: 217/D/5666$"
from Azarbiajan Shahid Madani university.
\nocite{*}

\bibliography{manuscript.bib}

\providecommand{\noopsort}[1]{}\providecommand{\singleletter}[1]{#1}%
\begin{thebibliography}{50}%
\makeatletter
\providecommand \@ifxundefined [1]{%
 \@ifx{#1\undefined}
}%
\providecommand \@ifnum [1]{%
 \ifnum #1\expandafter \@firstoftwo
 \else \expandafter \@secondoftwo
 \fi
}%
\providecommand \@ifx [1]{%
 \ifx #1\expandafter \@firstoftwo
 \else \expandafter \@secondoftwo
 \fi
}%
\providecommand \natexlab [1]{#1}%
\providecommand \enquote  [1]{``#1''}%
\providecommand \bibnamefont  [1]{#1}%
\providecommand \bibfnamefont [1]{#1}%
\providecommand \citenamefont [1]{#1}%
\providecommand \href@noop [0]{\@secondoftwo}%
\providecommand \href [0]{\begingroup \@sanitize@url \@href}%
\providecommand \@href[1]{\@@startlink{#1}\@@href}%
\providecommand \@@href[1]{\endgroup#1\@@endlink}%
\providecommand \@sanitize@url [0]{\catcode `\\12\catcode `\$12\catcode
  `\&12\catcode `\#12\catcode `\^12\catcode `\_12\catcode `\%12\relax}%
\providecommand \@@startlink[1]{}%
\providecommand \@@endlink[0]{}%
\providecommand \url  [0]{\begingroup\@sanitize@url \@url }%
\providecommand \@url [1]{\endgroup\@href {#1}{\urlprefix }}%
\providecommand \urlprefix  [0]{URL }%
\providecommand \Eprint [0]{\href }%
\providecommand \doibase [0]{http://dx.doi.org/}%
\providecommand \selectlanguage [0]{\@gobble}%
\providecommand \bibinfo  [0]{\@secondoftwo}%
\providecommand \bibfield  [0]{\@secondoftwo}%
\providecommand \translation [1]{[#1]}%
\providecommand \BibitemOpen [0]{}%
\providecommand \bibitemStop [0]{}%
\providecommand \bibitemNoStop [0]{.\EOS\space}%
\providecommand \EOS [0]{\spacefactor3000\relax}%
\providecommand \BibitemShut  [1]{\csname bibitem#1\endcsname}%
\let\auto@bib@innerbib\@empty
\bibitem [{\citenamefont {Zhong}\ \emph {et~al.}(2012)\citenamefont {Zhong},
  \citenamefont {Amorim}, \citenamefont {Scheicher}, \citenamefont {Pandey},\
  and\ \citenamefont {Karnac}}]{Zhong}%
  \BibitemOpen
  \bibfield  {author} {\bibinfo {author} {\bibfnamefont {X.}~\bibnamefont
  {Zhong}}, \bibinfo {author} {\bibfnamefont {R.~G.}\ \bibnamefont {Amorim}},
  \bibinfo {author} {\bibfnamefont {R.~H.}\ \bibnamefont {Scheicher}}, \bibinfo
  {author} {\bibfnamefont {R.}~\bibnamefont {Pandey}}, \ and\ \bibinfo {author}
  {\bibfnamefont {S.~P.}\ \bibnamefont {Karnac}},\ }\href@noop {} {\bibfield
  {journal} {\bibinfo  {journal} {Nanoscale}\ }\textbf {\bibinfo {volume}
  {4}},\ \bibinfo {pages} {5490} (\bibinfo {year} {2012})}\BibitemShut
  {NoStop}%
\bibitem [{\citenamefont {Savage}(2012)}]{Savage}%
  \BibitemOpen
  \bibfield  {author} {\bibinfo {author} {\bibfnamefont {N.}~\bibnamefont
  {Savage}},\ }\href@noop {} {\bibfield  {journal} {\bibinfo  {journal}
  {Materials science: super carbon, Nature}\ }\textbf {\bibinfo {volume}
  {483}},\ \bibinfo {pages} {S30} (\bibinfo {year} {2012})}\BibitemShut
  {NoStop}%
\bibitem [{\citenamefont {Conklin}\ \emph {et~al.}(2012)\citenamefont
  {Conklin}, \citenamefont {Nanayakkara}, \citenamefont {Park}, \citenamefont
  {Lagadec}, \citenamefont {Stecher}, \citenamefont {Therien},\ and\
  \citenamefont {Bonnell}}]{David}%
  \BibitemOpen
  \bibfield  {author} {\bibinfo {author} {\bibfnamefont {D.}~\bibnamefont
  {Conklin}}, \bibinfo {author} {\bibfnamefont {S.}~\bibnamefont
  {Nanayakkara}}, \bibinfo {author} {\bibfnamefont {T.-H.}\ \bibnamefont
  {Park}}, \bibinfo {author} {\bibfnamefont {M.~F.}\ \bibnamefont {Lagadec}},
  \bibinfo {author} {\bibfnamefont {J.~T.}\ \bibnamefont {Stecher}}, \bibinfo
  {author} {\bibfnamefont {M.~J.}\ \bibnamefont {Therien}}, \ and\ \bibinfo
  {author} {\bibfnamefont {D.~A.}\ \bibnamefont {Bonnell}},\ }\href@noop {}
  {\bibfield  {journal} {\bibinfo  {journal} {Nano Lett.}\ }\textbf {\bibinfo
  {volume} {12(5)}},\ \bibinfo {pages} {2414} (\bibinfo {year}
  {2012})}\BibitemShut {NoStop}%
\bibitem [{\citenamefont {Zhu}\ \emph {et~al.}(2012)\citenamefont {Zhu},
  \citenamefont {Low}, \citenamefont {Perebeinos}, \citenamefont {Bol},
  \citenamefont {Zhu}, \citenamefont {Yan}, \citenamefont {Tersoff},\ and\
  \citenamefont {Avouris}}]{Zhu}%
  \BibitemOpen
  \bibfield  {author} {\bibinfo {author} {\bibfnamefont {W.}~\bibnamefont
  {Zhu}}, \bibinfo {author} {\bibfnamefont {T.}~\bibnamefont {Low}}, \bibinfo
  {author} {\bibfnamefont {V.}~\bibnamefont {Perebeinos}}, \bibinfo {author}
  {\bibfnamefont {A.~A.}\ \bibnamefont {Bol}}, \bibinfo {author} {\bibfnamefont
  {Y.}~\bibnamefont {Zhu}}, \bibinfo {author} {\bibfnamefont {H.}~\bibnamefont
  {Yan}}, \bibinfo {author} {\bibfnamefont {J.}~\bibnamefont {Tersoff}}, \ and\
  \bibinfo {author} {\bibfnamefont {P.}~\bibnamefont {Avouris}},\ }\href@noop
  {} {\bibfield  {journal} {\bibinfo  {journal} {Nano Lett.}\ }\textbf
  {\bibinfo {volume} {12(7)}},\ \bibinfo {pages} {3431} (\bibinfo {year}
  {2012})}\BibitemShut {NoStop}%
\bibitem [{\citenamefont {Lherbier}\ \emph {et~al.}(2013)\citenamefont
  {Lherbier}, \citenamefont {Botello-Méndez},\ and\ \citenamefont
  {Charlier}}]{Lherbier}%
  \BibitemOpen
  \bibfield  {author} {\bibinfo {author} {\bibfnamefont {A.}~\bibnamefont
  {Lherbier}}, \bibinfo {author} {\bibfnamefont {A.~R.}\ \bibnamefont
  {Botello-Méndez}}, \ and\ \bibinfo {author} {\bibfnamefont {J.-C.}\
  \bibnamefont {Charlier}},\ }\href@noop {} {\bibfield  {journal} {\bibinfo
  {journal} {Nano Lett.}\ }\textbf {\bibinfo {volume} {13(4)}},\ \bibinfo
  {pages} {1446} (\bibinfo {year} {2013})}\BibitemShut {NoStop}%
\bibitem [{\citenamefont {Prasongkit}\ \emph {et~al.}(2013)\citenamefont
  {Prasongkit}, \citenamefont {Grigoriev},\ and\ \citenamefont
  {Ahuja}}]{Prasongkit}%
  \BibitemOpen
  \bibfield  {author} {\bibinfo {author} {\bibfnamefont {J.}~\bibnamefont
  {Prasongkit}}, \bibinfo {author} {\bibfnamefont {A.}~\bibnamefont
  {Grigoriev}}, \ and\ \bibinfo {author} {\bibfnamefont {R.}~\bibnamefont
  {Ahuja}},\ }\href@noop {} {\bibfield  {journal} {\bibinfo  {journal}
  {Phys.Rev. B}\ }\textbf {\bibinfo {volume} {87}},\ \bibinfo {pages} {155434}
  (\bibinfo {year} {2013})}\BibitemShut {NoStop}%
\bibitem [{\citenamefont {Gruznev}\ \emph {et~al.}(2015)\citenamefont
  {Gruznev}, \citenamefont {Bondarenko}, \citenamefont {Matetskiy},
  \citenamefont {Tupchaya}, \citenamefont {Alekseev}, \citenamefont {Hsing},
  \citenamefont {Wei}, \citenamefont {Eremeev}, \citenamefont {Zotov},\ and\
  \citenamefont {Saranin}}]{Gruznev}%
  \BibitemOpen
  \bibfield  {author} {\bibinfo {author} {\bibfnamefont {D.~V.}\ \bibnamefont
  {Gruznev}}, \bibinfo {author} {\bibfnamefont {L.~V.}\ \bibnamefont
  {Bondarenko}}, \bibinfo {author} {\bibfnamefont {A.~V.}\ \bibnamefont
  {Matetskiy}}, \bibinfo {author} {\bibfnamefont {A.~Y.}\ \bibnamefont
  {Tupchaya}}, \bibinfo {author} {\bibfnamefont {A.~A.}\ \bibnamefont
  {Alekseev}}, \bibinfo {author} {\bibfnamefont {C.~R.}\ \bibnamefont {Hsing}},
  \bibinfo {author} {\bibfnamefont {C.~M.}\ \bibnamefont {Wei}}, \bibinfo
  {author} {\bibfnamefont {S.~V.}\ \bibnamefont {Eremeev}}, \bibinfo {author}
  {\bibfnamefont {A.~V.}\ \bibnamefont {Zotov}}, \ and\ \bibinfo {author}
  {\bibfnamefont {A.~A.}\ \bibnamefont {Saranin}},\ }\href@noop {} {\bibfield
  {journal} {\bibinfo  {journal} {Phys. Rev. B}\ }\textbf {\bibinfo {volume}
  {91}},\ \bibinfo {pages} {035421} (\bibinfo {year} {2015})}\BibitemShut
  {NoStop}%
\bibitem [{\citenamefont {Benchamekh}\ \emph {et~al.}(2015)\citenamefont
  {Benchamekh}, \citenamefont {Raouafi}, \citenamefont {Even}, \citenamefont
  {Larbi}, \citenamefont {Voisin},\ and\ \citenamefont {Jancu}}]{Benchamekh}%
  \BibitemOpen
  \bibfield  {author} {\bibinfo {author} {\bibfnamefont {R.}~\bibnamefont
  {Benchamekh}}, \bibinfo {author} {\bibfnamefont {F.}~\bibnamefont {Raouafi}},
  \bibinfo {author} {\bibfnamefont {J.}~\bibnamefont {Even}}, \bibinfo {author}
  {\bibfnamefont {F.~B.~C.}\ \bibnamefont {Larbi}}, \bibinfo {author}
  {\bibfnamefont {P.}~\bibnamefont {Voisin}}, \ and\ \bibinfo {author}
  {\bibfnamefont {J.-M.}\ \bibnamefont {Jancu}},\ }\href@noop {} {\bibfield
  {journal} {\bibinfo  {journal} {Phys. Rev. B}\ }\textbf {\bibinfo {volume}
  {91}},\ \bibinfo {pages} {045118} (\bibinfo {year} {2015})}\BibitemShut
  {NoStop}%
\bibitem [{\citenamefont {Liu}\ \emph {et~al.}(2014)\citenamefont {Liu},
  \citenamefont {Wei}, \citenamefont {Yoon}, \citenamefont {Adak},
  \citenamefont {Ponce}, \citenamefont {Jiang}, \citenamefont {Jang},
  \citenamefont {Campos}, \citenamefont {Venkataraman},\ and\ \citenamefont
  {Neaton}}]{Zhen}%
  \BibitemOpen
  \bibfield  {author} {\bibinfo {author} {\bibfnamefont {Z.-F.}\ \bibnamefont
  {Liu}}, \bibinfo {author} {\bibfnamefont {S.}~\bibnamefont {Wei}}, \bibinfo
  {author} {\bibfnamefont {H.}~\bibnamefont {Yoon}}, \bibinfo {author}
  {\bibfnamefont {O.}~\bibnamefont {Adak}}, \bibinfo {author} {\bibfnamefont
  {I.}~\bibnamefont {Ponce}}, \bibinfo {author} {\bibfnamefont
  {Y.}~\bibnamefont {Jiang}}, \bibinfo {author} {\bibfnamefont {W.-D.}\
  \bibnamefont {Jang}}, \bibinfo {author} {\bibfnamefont {L.~M.}\ \bibnamefont
  {Campos}}, \bibinfo {author} {\bibfnamefont {L.}~\bibnamefont
  {Venkataraman}}, \ and\ \bibinfo {author} {\bibfnamefont {J.~B.}\
  \bibnamefont {Neaton}},\ }\href@noop {} {\bibfield  {journal} {\bibinfo
  {journal} {Nano. Lett.}\ }\textbf {\bibinfo {volume} {14(9)}},\ \bibinfo
  {pages} {5365} (\bibinfo {year} {2014})}\BibitemShut {NoStop}%
\bibitem [{\citenamefont {Geim}\ and\ \citenamefont {Novoselove}(2007)}]{Geim}%
  \BibitemOpen
  \bibfield  {author} {\bibinfo {author} {\bibfnamefont {A.~K.}\ \bibnamefont
  {Geim}}\ and\ \bibinfo {author} {\bibfnamefont {K.}~\bibnamefont
  {Novoselove}},\ }\href@noop {} {\bibfield  {journal} {\bibinfo  {journal}
  {Nature Materials}\ }\textbf {\bibinfo {volume} {6}},\ \bibinfo {pages} {183}
  (\bibinfo {year} {2007})}\BibitemShut {NoStop}%
\bibitem [{\citenamefont {Novoselov}\ \emph {et~al.}(2004)\citenamefont
  {Novoselov}, \citenamefont {Geim}, \citenamefont {Morozov}, \citenamefont
  {Jiang}, \citenamefont {Zhang}, \citenamefont {Dubonos}, \citenamefont
  {Grigorieva},\ and\ \citenamefont {Firsov}}]{Novoselov}%
  \BibitemOpen
  \bibfield  {author} {\bibinfo {author} {\bibfnamefont {K.~S.}\ \bibnamefont
  {Novoselov}}, \bibinfo {author} {\bibfnamefont {A.~K.}\ \bibnamefont {Geim}},
  \bibinfo {author} {\bibfnamefont {S.~V.}\ \bibnamefont {Morozov}}, \bibinfo
  {author} {\bibfnamefont {D.}~\bibnamefont {Jiang}}, \bibinfo {author}
  {\bibfnamefont {Y.}~\bibnamefont {Zhang}}, \bibinfo {author} {\bibfnamefont
  {S.~V.}\ \bibnamefont {Dubonos}}, \bibinfo {author} {\bibfnamefont {I.~V.}\
  \bibnamefont {Grigorieva}}, \ and\ \bibinfo {author} {\bibfnamefont {A.~A.}\
  \bibnamefont {Firsov}},\ }\href@noop {} {\bibfield  {journal} {\bibinfo
  {journal} {Science}\ }\textbf {\bibinfo {volume} {306}},\ \bibinfo {pages}
  {666} (\bibinfo {year} {2004})}\BibitemShut {NoStop}%
\bibitem [{\citenamefont {Motta}\ \emph {et~al.}(2012)\citenamefont {Motta},
  \citenamefont {Sanchez-Portal},\ and\ \citenamefont {Trioni}}]{Motta}%
  \BibitemOpen
  \bibfield  {author} {\bibinfo {author} {\bibfnamefont {C.}~\bibnamefont
  {Motta}}, \bibinfo {author} {\bibfnamefont {D.}~\bibnamefont
  {Sanchez-Portal}}, \ and\ \bibinfo {author} {\bibfnamefont {M.~I.}\
  \bibnamefont {Trioni}},\ }\href@noop {} {\bibfield  {journal} {\bibinfo
  {journal} {Phys. Chem. Chem. Phys.}\ }\textbf {\bibinfo {volume} {14}},\
  \bibinfo {pages} {10683} (\bibinfo {year} {2012})}\BibitemShut {NoStop}%
\bibitem [{\citenamefont {Lherbier}\ \emph {et~al.}(2014)\citenamefont
  {Lherbier}, \citenamefont {Terrones},\ and\ \citenamefont
  {Charlier}}]{Lherbier1}%
  \BibitemOpen
  \bibfield  {author} {\bibinfo {author} {\bibfnamefont {A.}~\bibnamefont
  {Lherbier}}, \bibinfo {author} {\bibfnamefont {H.}~\bibnamefont {Terrones}},
  \ and\ \bibinfo {author} {\bibfnamefont {J.-C.}\ \bibnamefont {Charlier}},\
  }\href@noop {} {\bibfield  {journal} {\bibinfo  {journal} {Phys. Rev. B}\
  }\textbf {\bibinfo {volume} {90}},\ \bibinfo {pages} {125434} (\bibinfo
  {year} {2014})}\BibitemShut {NoStop}%
\bibitem [{\citenamefont {Charlier}\ \emph {et~al.}(1991)\citenamefont
  {Charlier}, \citenamefont {Michenaud}, \citenamefont {Gonze},\ and\
  \citenamefont {Vigneron}}]{Charlier}%
  \BibitemOpen
  \bibfield  {author} {\bibinfo {author} {\bibfnamefont {J.-C.}\ \bibnamefont
  {Charlier}}, \bibinfo {author} {\bibfnamefont {J.-P.}\ \bibnamefont
  {Michenaud}}, \bibinfo {author} {\bibfnamefont {X.}~\bibnamefont {Gonze}}, \
  and\ \bibinfo {author} {\bibfnamefont {J.-P.}\ \bibnamefont {Vigneron}},\
  }\href@noop {} {\bibfield  {journal} {\bibinfo  {journal} {Phys. Rev. B}\
  }\textbf {\bibinfo {volume} {44}},\ \bibinfo {pages} {13237} (\bibinfo {year}
  {1991})}\BibitemShut {NoStop}%
\bibitem [{\citenamefont {Neto}\ \emph {et~al.}(2009)\citenamefont {Neto},
  \citenamefont {Guinea}, \citenamefont {Peres}, \citenamefont {Novoselov},\
  and\ \citenamefont {Geim}}]{Castro}%
  \BibitemOpen
  \bibfield  {author} {\bibinfo {author} {\bibfnamefont {A.~H.~C.}\
  \bibnamefont {Neto}}, \bibinfo {author} {\bibfnamefont {F.}~\bibnamefont
  {Guinea}}, \bibinfo {author} {\bibfnamefont {N.~M.~R.}\ \bibnamefont
  {Peres}}, \bibinfo {author} {\bibfnamefont {K.~S.}\ \bibnamefont
  {Novoselov}}, \ and\ \bibinfo {author} {\bibfnamefont {A.~K.}\ \bibnamefont
  {Geim}},\ }\href@noop {} {\bibfield  {journal} {\bibinfo  {journal} {Rev.
  Mod. Phys.}\ }\textbf {\bibinfo {volume} {81}},\ \bibinfo {pages} {109}
  (\bibinfo {year} {2009})}\BibitemShut {NoStop}%
\bibitem [{\citenamefont {Neto}\ \emph {et~al.}(1947)\citenamefont {Neto},
  \citenamefont {Guinea}, \citenamefont {Peres}, \citenamefont {Novoselov},\
  and\ \citenamefont {Geim}}]{Wallace}%
  \BibitemOpen
  \bibfield  {author} {\bibinfo {author} {\bibfnamefont {A.~H.~C.}\
  \bibnamefont {Neto}}, \bibinfo {author} {\bibfnamefont {F.}~\bibnamefont
  {Guinea}}, \bibinfo {author} {\bibfnamefont {N.~M.~R.}\ \bibnamefont
  {Peres}}, \bibinfo {author} {\bibfnamefont {K.~S.}\ \bibnamefont
  {Novoselov}}, \ and\ \bibinfo {author} {\bibfnamefont {A.~K.}\ \bibnamefont
  {Geim}},\ }\href@noop {} {\bibfield  {journal} {\bibinfo  {journal} {hys.
  Rev.}\ }\textbf {\bibinfo {volume} {71}},\ \bibinfo {pages} {622} (\bibinfo
  {year} {1947})}\BibitemShut {NoStop}%
\bibitem [{\citenamefont {Torres}\ \emph {et~al.}(2014)\citenamefont {Torres},
  \citenamefont {Roche},\ and\ \citenamefont {Charlier}}]{Luis}%
  \BibitemOpen
  \bibfield  {author} {\bibinfo {author} {\bibfnamefont {L.~E. F.~F.}\
  \bibnamefont {Torres}}, \bibinfo {author} {\bibfnamefont {S.}~\bibnamefont
  {Roche}}, \ and\ \bibinfo {author} {\bibfnamefont {J.-C.}\ \bibnamefont
  {Charlier}},\ }\href@noop {} {\emph {\bibinfo {title} {Introduction to
  Graphene-Based Nanomaterials: From Electronic Structure to quantum
  transport}}}\ (\bibinfo  {publisher} {Cambridge University Press},\ \bibinfo
  {year} {2014})\BibitemShut {NoStop}%
\bibitem [{\citenamefont {Wakabayashi}\ \emph {et~al.}(2009)\citenamefont
  {Wakabayashi}, \citenamefont {Takane}, \citenamefont {Yamamoto},\ and\
  \citenamefont {Sigrist}}]{Wakabayashi}%
  \BibitemOpen
  \bibfield  {author} {\bibinfo {author} {\bibfnamefont {K.}~\bibnamefont
  {Wakabayashi}}, \bibinfo {author} {\bibfnamefont {Y.}~\bibnamefont {Takane}},
  \bibinfo {author} {\bibfnamefont {M.}~\bibnamefont {Yamamoto}}, \ and\
  \bibinfo {author} {\bibfnamefont {M.}~\bibnamefont {Sigrist}},\ }\href@noop
  {} {\bibfield  {journal} {\bibinfo  {journal} {New J. Phys.}\ }\textbf
  {\bibinfo {volume} {11}},\ \bibinfo {pages} {095016} (\bibinfo {year}
  {2009})}\BibitemShut {NoStop}%
\bibitem [{\citenamefont {Avouris}\ \emph {et~al.}(2007)\citenamefont
  {Avouris}, \citenamefont {Chen},\ and\ \citenamefont {Perebeinos}}]{Avouris}%
  \BibitemOpen
  \bibfield  {author} {\bibinfo {author} {\bibfnamefont {P.}~\bibnamefont
  {Avouris}}, \bibinfo {author} {\bibfnamefont {Z.}~\bibnamefont {Chen}}, \
  and\ \bibinfo {author} {\bibfnamefont {V.}~\bibnamefont {Perebeinos}},\
  }\href@noop {} {\bibfield  {journal} {\bibinfo  {journal} {Nature
  Nanotechnology}\ }\textbf {\bibinfo {volume} {2}},\ \bibinfo {pages} {605}
  (\bibinfo {year} {2007})}\BibitemShut {NoStop}%
\bibitem [{\citenamefont {Wang}\ \emph {et~al.}(2010)\citenamefont {Wang},
  \citenamefont {Ang}, \citenamefont {Wang}, \citenamefont {Tang},
  \citenamefont {Thong},\ and\ \citenamefont {Loh}}]{Wang}%
  \BibitemOpen
  \bibfield  {author} {\bibinfo {author} {\bibfnamefont {S.}~\bibnamefont
  {Wang}}, \bibinfo {author} {\bibfnamefont {P.~K.}\ \bibnamefont {Ang}},
  \bibinfo {author} {\bibfnamefont {Z.}~\bibnamefont {Wang}}, \bibinfo {author}
  {\bibfnamefont {A.~L.~L.}\ \bibnamefont {Tang}}, \bibinfo {author}
  {\bibfnamefont {J.~T.~L.}\ \bibnamefont {Thong}}, \ and\ \bibinfo {author}
  {\bibfnamefont {K.~P.}\ \bibnamefont {Loh}},\ }\href@noop {} {\bibfield
  {journal} {\bibinfo  {journal} {Nano Lett.}\ }\textbf {\bibinfo {volume}
  {10(1)}},\ \bibinfo {pages} {92} (\bibinfo {year} {2010})}\BibitemShut
  {NoStop}%
\bibitem [{\citenamefont {Kim}\ \emph {et~al.}(2013)\citenamefont {Kim},
  \citenamefont {Shin}, \citenamefont {Kim}, \citenamefont {Kang},
  \citenamefont {Kim}, \citenamefont {Jang}, \citenamefont {Joo}, \citenamefont
  {Lee}, \citenamefont {Kim}, \citenamefont {Choi},\ and\ \citenamefont
  {Hwang}}]{Sung}%
  \BibitemOpen
  \bibfield  {author} {\bibinfo {author} {\bibfnamefont {S.}~\bibnamefont
  {Kim}}, \bibinfo {author} {\bibfnamefont {D.~H.}\ \bibnamefont {Shin}},
  \bibinfo {author} {\bibfnamefont {C.~O.}\ \bibnamefont {Kim}}, \bibinfo
  {author} {\bibfnamefont {S.~S.}\ \bibnamefont {Kang}}, \bibinfo {author}
  {\bibfnamefont {J.~M.}\ \bibnamefont {Kim}}, \bibinfo {author} {\bibfnamefont
  {C.~W.}\ \bibnamefont {Jang}}, \bibinfo {author} {\bibfnamefont {S.~S.}\
  \bibnamefont {Joo}}, \bibinfo {author} {\bibfnamefont {J.~S.}\ \bibnamefont
  {Lee}}, \bibinfo {author} {\bibfnamefont {J.~H.}\ \bibnamefont {Kim}},
  \bibinfo {author} {\bibfnamefont {S.-H.}\ \bibnamefont {Choi}}, \ and\
  \bibinfo {author} {\bibfnamefont {E.}~\bibnamefont {Hwang}},\ }\href@noop {}
  {\bibfield  {journal} {\bibinfo  {journal} {ACS Nano}\ }\textbf {\bibinfo
  {volume} {7(6)}},\ \bibinfo {pages} {5168} (\bibinfo {year}
  {2013})}\BibitemShut {NoStop}%
\bibitem [{\citenamefont {Morozov}\ \emph {et~al.}(2006)\citenamefont
  {Morozov}, \citenamefont {Novoselov}, \citenamefont {Katsnelson},
  \citenamefont {Schedin}, \citenamefont {Ponomarenko}, \citenamefont {Jiang},\
  and\ \citenamefont {Geim}}]{Morozov}%
  \BibitemOpen
  \bibfield  {author} {\bibinfo {author} {\bibfnamefont {S.~V.}\ \bibnamefont
  {Morozov}}, \bibinfo {author} {\bibfnamefont {K.~S.}\ \bibnamefont
  {Novoselov}}, \bibinfo {author} {\bibfnamefont {M.~I.}\ \bibnamefont
  {Katsnelson}}, \bibinfo {author} {\bibfnamefont {F.}~\bibnamefont {Schedin}},
  \bibinfo {author} {\bibfnamefont {L.~A.}\ \bibnamefont {Ponomarenko}},
  \bibinfo {author} {\bibfnamefont {D.}~\bibnamefont {Jiang}}, \ and\ \bibinfo
  {author} {\bibfnamefont {A.~K.}\ \bibnamefont {Geim}},\ }\href@noop {}
  {\bibfield  {journal} {\bibinfo  {journal} {Phys. Rev. Lett.}\ }\textbf
  {\bibinfo {volume} {97}},\ \bibinfo {pages} {016801} (\bibinfo {year}
  {2006})}\BibitemShut {NoStop}%
\bibitem [{\citenamefont {Recher}\ \emph {et~al.}(2007)\citenamefont {Recher},
  \citenamefont {Trauzettel}, \citenamefont {Rycerz}, \citenamefont {Blanter},
  \citenamefont {Beenakker},\ and\ \citenamefont {Morpurgo}}]{Recher}%
  \BibitemOpen
  \bibfield  {author} {\bibinfo {author} {\bibfnamefont {P.}~\bibnamefont
  {Recher}}, \bibinfo {author} {\bibfnamefont {B.}~\bibnamefont {Trauzettel}},
  \bibinfo {author} {\bibfnamefont {A.}~\bibnamefont {Rycerz}}, \bibinfo
  {author} {\bibfnamefont {Y.~M.}\ \bibnamefont {Blanter}}, \bibinfo {author}
  {\bibfnamefont {C.~W.~J.}\ \bibnamefont {Beenakker}}, \ and\ \bibinfo
  {author} {\bibfnamefont {A.~F.}\ \bibnamefont {Morpurgo}},\ }\href@noop {}
  {\bibfield  {journal} {\bibinfo  {journal} {Phys. Rev. B}\ }\textbf {\bibinfo
  {volume} {76}},\ \bibinfo {pages} {235404} (\bibinfo {year}
  {2007})}\BibitemShut {NoStop}%
\bibitem [{\citenamefont {Russo}\ \emph {et~al.}(2008)\citenamefont {Russo},
  \citenamefont {Oostinga}, \citenamefont {Wehenkel}, \citenamefont {Heersche},
  \citenamefont {Sobhani}, \citenamefont {Vandersypen},\ and\ \citenamefont
  {Morpurgo}}]{Russo}%
  \BibitemOpen
  \bibfield  {author} {\bibinfo {author} {\bibfnamefont {S.}~\bibnamefont
  {Russo}}, \bibinfo {author} {\bibfnamefont {J.~B.}\ \bibnamefont {Oostinga}},
  \bibinfo {author} {\bibfnamefont {D.}~\bibnamefont {Wehenkel}}, \bibinfo
  {author} {\bibfnamefont {H.~B.}\ \bibnamefont {Heersche}}, \bibinfo {author}
  {\bibfnamefont {S.~S.}\ \bibnamefont {Sobhani}}, \bibinfo {author}
  {\bibfnamefont {L.~M.~K.}\ \bibnamefont {Vandersypen}}, \ and\ \bibinfo
  {author} {\bibfnamefont {A.~F.}\ \bibnamefont {Morpurgo}},\ }\href@noop {}
  {\bibfield  {journal} {\bibinfo  {journal} {Phys. Rev. B}\ }\textbf {\bibinfo
  {volume} {77}},\ \bibinfo {pages} {085413} (\bibinfo {year}
  {2008})}\BibitemShut {NoStop}%
\bibitem [{\citenamefont {Yan}\ \emph {et~al.}(2007)\citenamefont {Yan},
  \citenamefont {Huang}, \citenamefont {Yu}, \citenamefont {Zheng},
  \citenamefont {Zang}, \citenamefont {Wu}, \citenamefont {Gu}, \citenamefont
  {Liu},\ and\ \citenamefont {Duan}}]{Yan}%
  \BibitemOpen
  \bibfield  {author} {\bibinfo {author} {\bibfnamefont {Q.}~\bibnamefont
  {Yan}}, \bibinfo {author} {\bibfnamefont {B.}~\bibnamefont {Huang}}, \bibinfo
  {author} {\bibfnamefont {J.}~\bibnamefont {Yu}}, \bibinfo {author}
  {\bibfnamefont {F.}~\bibnamefont {Zheng}}, \bibinfo {author} {\bibfnamefont
  {J.}~\bibnamefont {Zang}}, \bibinfo {author} {\bibfnamefont {J.}~\bibnamefont
  {Wu}}, \bibinfo {author} {\bibfnamefont {B.-L.}\ \bibnamefont {Gu}}, \bibinfo
  {author} {\bibfnamefont {F.}~\bibnamefont {Liu}}, \ and\ \bibinfo {author}
  {\bibfnamefont {W.}~\bibnamefont {Duan}},\ }\href@noop {} {\bibfield
  {journal} {\bibinfo  {journal} {Nano Lett.}\ }\textbf {\bibinfo {volume}
  {7(6)}},\ \bibinfo {pages} {1469} (\bibinfo {year} {2007})}\BibitemShut
  {NoStop}%
\bibitem [{\citenamefont {Riedl}\ \emph {et~al.}(2010)\citenamefont {Riedl},
  \citenamefont {Coletti},\ and\ \citenamefont {Starke}}]{Riedl}%
  \BibitemOpen
  \bibfield  {author} {\bibinfo {author} {\bibfnamefont {C.}~\bibnamefont
  {Riedl}}, \bibinfo {author} {\bibfnamefont {C.}~\bibnamefont {Coletti}}, \
  and\ \bibinfo {author} {\bibfnamefont {U.}~\bibnamefont {Starke}},\
  }\href@noop {} {\bibfield  {journal} {\bibinfo  {journal} {J. Phys. D: Appl.
  Phys.}\ }\textbf {\bibinfo {volume} {43}},\ \bibinfo {pages} {374009}
  (\bibinfo {year} {2010})}\BibitemShut {NoStop}%
\bibitem [{\citenamefont {Kaukonen}\ \emph {et~al.}(2013)\citenamefont
  {Kaukonen}, \citenamefont {Krasheninnikov}, \citenamefont {Kauppinen},\ and\
  \citenamefont {Nieminen}}]{Kaukonen}%
  \BibitemOpen
  \bibfield  {author} {\bibinfo {author} {\bibfnamefont {M.}~\bibnamefont
  {Kaukonen}}, \bibinfo {author} {\bibfnamefont {A.~V.}\ \bibnamefont
  {Krasheninnikov}}, \bibinfo {author} {\bibfnamefont {E.}~\bibnamefont
  {Kauppinen}}, \ and\ \bibinfo {author} {\bibfnamefont {R.~M.}\ \bibnamefont
  {Nieminen}},\ }\href@noop {} {\bibfield  {journal} {\bibinfo  {journal} {ACS
  Catal.}\ }\textbf {\bibinfo {volume} {3(2)}},\ \bibinfo {pages} {159}
  (\bibinfo {year} {2013})}\BibitemShut {NoStop}%
\bibitem [{\citenamefont {Plachinda}\ \emph {et~al.}(2011)\citenamefont
  {Plachinda}, \citenamefont {Evans},\ and\ \citenamefont {Solanki}}]{Paul}%
  \BibitemOpen
  \bibfield  {author} {\bibinfo {author} {\bibfnamefont {P.}~\bibnamefont
  {Plachinda}}, \bibinfo {author} {\bibfnamefont {D.~R.}\ \bibnamefont
  {Evans}}, \ and\ \bibinfo {author} {\bibfnamefont {R.}~\bibnamefont
  {Solanki}},\ }\href@noop {} {\bibfield  {journal} {\bibinfo  {journal} {J.
  Chem. Phys.}\ }\textbf {\bibinfo {volume} {135}},\ \bibinfo {pages} {044103}
  (\bibinfo {year} {2011})}\BibitemShut {NoStop}%
\bibitem [{\citenamefont {Yang}\ \emph {et~al.}(2013)\citenamefont {Yang},
  \citenamefont {Feng}, \citenamefont {Hong}, \citenamefont {Cai},\ and\
  \citenamefont {Liu}}]{Yang}%
  \BibitemOpen
  \bibfield  {author} {\bibinfo {author} {\bibfnamefont {K.}~\bibnamefont
  {Yang}}, \bibinfo {author} {\bibfnamefont {L.}~\bibnamefont {Feng}}, \bibinfo
  {author} {\bibfnamefont {H.}~\bibnamefont {Hong}}, \bibinfo {author}
  {\bibfnamefont {W.}~\bibnamefont {Cai}}, \ and\ \bibinfo {author}
  {\bibfnamefont {Z.}~\bibnamefont {Liu}},\ }\href@noop {} {\bibfield
  {journal} {\bibinfo  {journal} {Nature Protocols}\ }\textbf {\bibinfo
  {volume} {8}},\ \bibinfo {pages} {2392} (\bibinfo {year} {2013})}\BibitemShut
  {NoStop}%
\bibitem [{\citenamefont {Wood}\ \emph {et~al.}(2012)\citenamefont {Wood},
  \citenamefont {Bhide}, \citenamefont {Dutta}, \citenamefont {Kandagal},
  \citenamefont {Pathak}, \citenamefont {Punnathanam}, \citenamefont {Ayappa},\
  and\ \citenamefont {Narasimhan}}]{Wood}%
  \BibitemOpen
  \bibfield  {author} {\bibinfo {author} {\bibfnamefont {B.~C.}\ \bibnamefont
  {Wood}}, \bibinfo {author} {\bibfnamefont {S.~Y.}\ \bibnamefont {Bhide}},
  \bibinfo {author} {\bibfnamefont {D.}~\bibnamefont {Dutta}}, \bibinfo
  {author} {\bibfnamefont {V.~S.}\ \bibnamefont {Kandagal}}, \bibinfo {author}
  {\bibfnamefont {A.~D.}\ \bibnamefont {Pathak}}, \bibinfo {author}
  {\bibfnamefont {S.~N.}\ \bibnamefont {Punnathanam}}, \bibinfo {author}
  {\bibfnamefont {K.~G.}\ \bibnamefont {Ayappa}}, \ and\ \bibinfo {author}
  {\bibfnamefont {S.}~\bibnamefont {Narasimhan}},\ }\href@noop {} {\bibfield
  {journal} {\bibinfo  {journal} {J. Chem. Phys.}\ }\textbf {\bibinfo {volume}
  {137}},\ \bibinfo {pages} {0547702} (\bibinfo {year} {2012})}\BibitemShut
  {NoStop}%
\bibitem [{\citenamefont {Lopez-Bezanilla}\ \emph {et~al.}(2009)\citenamefont
  {Lopez-Bezanilla}, \citenamefont {Triozon},\ and\ \citenamefont
  {Roche}}]{Bezanilla}%
  \BibitemOpen
  \bibfield  {author} {\bibinfo {author} {\bibfnamefont {A.}~\bibnamefont
  {Lopez-Bezanilla}}, \bibinfo {author} {\bibfnamefont {F.}~\bibnamefont
  {Triozon}}, \ and\ \bibinfo {author} {\bibfnamefont {S.}~\bibnamefont
  {Roche}},\ }\href@noop {} {\bibfield  {journal} {\bibinfo  {journal} {Nano
  Lett.}\ }\textbf {\bibinfo {volume} {9(7)}},\ \bibinfo {pages} {2537}
  (\bibinfo {year} {2009})}\BibitemShut {NoStop}%
\bibitem [{\citenamefont {Yamacli}(2014)}]{Yamacli}%
  \BibitemOpen
  \bibfield  {author} {\bibinfo {author} {\bibfnamefont {S.}~\bibnamefont
  {Yamacli}},\ }\href@noop {} {\bibfield  {journal} {\bibinfo  {journal}
  {Nano-Micro Letters}\ }\textbf {\bibinfo {volume} {7(1)}},\  (\bibinfo {year}
  {2014})}\BibitemShut {NoStop}%
\bibitem [{\citenamefont {Beneden}\ and\ \citenamefont
  {Bruno}(2003)}]{Beneden}%
  \BibitemOpen
  \bibfield  {author} {\bibinfo {author} {\bibfnamefont {V.}~\bibnamefont
  {Beneden}}\ and\ \bibinfo {author} {\bibnamefont {Bruno}},\ }\href@noop {}
  {\bibfield  {journal} {\bibinfo  {journal} {Power Electronics Technology}\
  }\textbf {\bibinfo {volume} {29}},\ \bibinfo {pages} {26} (\bibinfo {year}
  {2003})}\BibitemShut {NoStop}%
\bibitem [{\citenamefont {Liu}\ \emph {et~al.}(2012)\citenamefont {Liu},
  \citenamefont {Yao}, \citenamefont {Zhou}, \citenamefont {Qin},\ and\
  \citenamefont {Liu}}]{Liu}%
  \BibitemOpen
  \bibfield  {author} {\bibinfo {author} {\bibfnamefont {Q.}~\bibnamefont
  {Liu}}, \bibinfo {author} {\bibfnamefont {X.}~\bibnamefont {Yao}}, \bibinfo
  {author} {\bibfnamefont {X.}~\bibnamefont {Zhou}}, \bibinfo {author}
  {\bibfnamefont {Z.}~\bibnamefont {Qin}}, \ and\ \bibinfo {author}
  {\bibfnamefont {Z.}~\bibnamefont {Liu}},\ }\href@noop {} {\bibfield
  {journal} {\bibinfo  {journal} {Scripta Materialia}\ }\textbf {\bibinfo
  {volume} {66}},\ \bibinfo {pages} {113} (\bibinfo {year} {2012})}\BibitemShut
  {NoStop}%
\bibitem [{\citenamefont {Lin}\ \emph {et~al.}(2007)\citenamefont {Lin},
  \citenamefont {Lu},\ and\ \citenamefont {Chen}}]{Lin}%
  \BibitemOpen
  \bibfield  {author} {\bibinfo {author} {\bibfnamefont {H.}~\bibnamefont
  {Lin}}, \bibinfo {author} {\bibfnamefont {W.}~\bibnamefont {Lu}}, \ and\
  \bibinfo {author} {\bibfnamefont {G.}~\bibnamefont {Chen}},\ }\href@noop {}
  {\bibfield  {journal} {\bibinfo  {journal} {Physica B: Condensed Matter}\
  }\textbf {\bibinfo {volume} {400}},\ \bibinfo {pages} {229} (\bibinfo {year}
  {2007})}\BibitemShut {NoStop}%
\bibitem [{\citenamefont {Bidadi}\ \emph {et~al.}(2013)\citenamefont {Bidadi},
  \citenamefont {Olad}, \citenamefont {Parhizkar}, \citenamefont {Aref},\ and\
  \citenamefont {Ghafouri}}]{Bidadi}%
  \BibitemOpen
  \bibfield  {author} {\bibinfo {author} {\bibfnamefont {H.}~\bibnamefont
  {Bidadi}}, \bibinfo {author} {\bibfnamefont {A.}~\bibnamefont {Olad}},
  \bibinfo {author} {\bibfnamefont {M.}~\bibnamefont {Parhizkar}}, \bibinfo
  {author} {\bibfnamefont {S.~M.}\ \bibnamefont {Aref}}, \ and\ \bibinfo
  {author} {\bibfnamefont {M.}~\bibnamefont {Ghafouri}},\ }\href@noop {}
  {\bibfield  {journal} {\bibinfo  {journal} {Vacuum}\ }\textbf {\bibinfo
  {volume} {87}},\ \bibinfo {pages} {50} (\bibinfo {year} {2013})}\BibitemShut
  {NoStop}%
\bibitem [{\citenamefont {Datta}(1995)}]{Datta}%
  \BibitemOpen
  \bibfield  {author} {\bibinfo {author} {\bibfnamefont {S.}~\bibnamefont
  {Datta}},\ }\href@noop {} {\emph {\bibinfo {title} {Electronic Transport in
  Mesoscopic Systems}}}\ (\bibinfo  {publisher} {Cambridge University Press},\
  \bibinfo {year} {1995})\BibitemShut {NoStop}%
\bibitem [{\citenamefont {Ozaki}\ \emph {et~al.}(2010)\citenamefont {Ozaki},
  \citenamefont {Nishio},\ and\ \citenamefont {Kino}}]{Ozaki5}%
  \BibitemOpen
  \bibfield  {author} {\bibinfo {author} {\bibfnamefont {T.}~\bibnamefont
  {Ozaki}}, \bibinfo {author} {\bibfnamefont {K.}~\bibnamefont {Nishio}}, \
  and\ \bibinfo {author} {\bibfnamefont {H.}~\bibnamefont {Kino}},\ }\href@noop
  {} {\bibfield  {journal} {\bibinfo  {journal} {Phys. Rev. B}\ }\textbf
  {\bibinfo {volume} {81}},\ \bibinfo {pages} {035116} (\bibinfo {year}
  {2010})}\BibitemShut {NoStop}%
\bibitem [{\citenamefont {Liang}\ \emph {et~al.}(2004)\citenamefont {Liang},
  \citenamefont {Ghosh}, \citenamefont {Paulsson},\ and\ \citenamefont
  {Datta}}]{Liang}%
  \BibitemOpen
  \bibfield  {author} {\bibinfo {author} {\bibfnamefont {G.~C.}\ \bibnamefont
  {Liang}}, \bibinfo {author} {\bibfnamefont {A.~W.}\ \bibnamefont {Ghosh}},
  \bibinfo {author} {\bibfnamefont {M.}~\bibnamefont {Paulsson}}, \ and\
  \bibinfo {author} {\bibfnamefont {S.}~\bibnamefont {Datta}},\ }\href@noop {}
  {\bibfield  {journal} {\bibinfo  {journal} {Phys. Rev. B}\ }\textbf {\bibinfo
  {volume} {69}},\ \bibinfo {pages} {115302} (\bibinfo {year}
  {2004})}\BibitemShut {NoStop}%
\bibitem [{\citenamefont {Landauer}(1957)}]{Landauer}%
  \BibitemOpen
  \bibfield  {author} {\bibinfo {author} {\bibfnamefont {R.}~\bibnamefont
  {Landauer}},\ }\href@noop {} {\bibfield  {journal} {\bibinfo  {journal} {IBM
  J. Res. Dev.}\ }\textbf {\bibinfo {volume} {1}},\ \bibinfo {pages} {223}
  (\bibinfo {year} {1957})}\BibitemShut {NoStop}%
\bibitem [{\citenamefont {Büttiker}(1986)}]{Buttiker}%
  \BibitemOpen
  \bibfield  {author} {\bibinfo {author} {\bibfnamefont {M.}~\bibnamefont
  {Büttiker}},\ }\href@noop {} {\bibfield  {journal} {\bibinfo  {journal}
  {Phys. Rev. Lett.}\ }\textbf {\bibinfo {volume} {57}},\ \bibinfo {pages}
  {1761} (\bibinfo {year} {1986})}\BibitemShut {NoStop}%
\bibitem [{OPE()}]{OPENMX}%
  \BibitemOpen
  \href {http://www.openmx-square.org} {}\bibinfo {note} {The code, OPENMX,
  pseudoatomic basis functions, and pseudo- potentials are available on a web
  site: http://www.openmx-square.org}\BibitemShut {NoStop}%
\bibitem [{\citenamefont {Ozaki}\ and\ \citenamefont
  {Kino}(2004{\natexlab{a}})}]{Ozaki3}%
  \BibitemOpen
  \bibfield  {author} {\bibinfo {author} {\bibfnamefont {T.}~\bibnamefont
  {Ozaki}}\ and\ \bibinfo {author} {\bibfnamefont {H.}~\bibnamefont {Kino}},\
  }\href@noop {} {\bibfield  {journal} {\bibinfo  {journal} {J. Chem. Phys.}\
  }\textbf {\bibinfo {volume} {121}},\ \bibinfo {pages} {10879} (\bibinfo
  {year} {2004}{\natexlab{a}})}\BibitemShut {NoStop}%
\bibitem [{\citenamefont {Ozaki}(2003)}]{Ozaki6}%
  \BibitemOpen
  \bibfield  {author} {\bibinfo {author} {\bibfnamefont {T.}~\bibnamefont
  {Ozaki}},\ }\href@noop {} {\bibfield  {journal} {\bibinfo  {journal} {Phys.
  Rev. B}\ }\textbf {\bibinfo {volume} {67}},\ \bibinfo {pages} {155108}
  (\bibinfo {year} {2003})}\BibitemShut {NoStop}%
\bibitem [{\citenamefont {Ozaki}\ and\ \citenamefont
  {Kino}(2004{\natexlab{b}})}]{Ozaki2}%
  \BibitemOpen
  \bibfield  {author} {\bibinfo {author} {\bibfnamefont {T.}~\bibnamefont
  {Ozaki}}\ and\ \bibinfo {author} {\bibfnamefont {H.}~\bibnamefont {Kino}},\
  }\href@noop {} {\bibfield  {journal} {\bibinfo  {journal} {Phys. Rev. B}\
  }\textbf {\bibinfo {volume} {69}},\ \bibinfo {pages} {195113} (\bibinfo
  {year} {2004}{\natexlab{b}})}\BibitemShut {NoStop}%
\bibitem [{\citenamefont {Ceperley}\ and\ \citenamefont
  {Alder}(1980)}]{Ceperley}%
  \BibitemOpen
  \bibfield  {author} {\bibinfo {author} {\bibfnamefont {D.~M.}\ \bibnamefont
  {Ceperley}}\ and\ \bibinfo {author} {\bibfnamefont {B.~J.}\ \bibnamefont
  {Alder}},\ }\href@noop {} {\bibfield  {journal} {\bibinfo  {journal} {Phys.
  Rev. Lett.}\ }\textbf {\bibinfo {volume} {45}},\ \bibinfo {pages} {566}
  (\bibinfo {year} {1980})}\BibitemShut {NoStop}%
\bibitem [{\citenamefont {Perdew}\ and\ \citenamefont {Zunger}(1981)}]{Perdew}%
  \BibitemOpen
  \bibfield  {author} {\bibinfo {author} {\bibfnamefont {J.~P.}\ \bibnamefont
  {Perdew}}\ and\ \bibinfo {author} {\bibfnamefont {A.}~\bibnamefont
  {Zunger}},\ }\href@noop {} {\bibfield  {journal} {\bibinfo  {journal} {Phys.
  Rev. B}\ }\textbf {\bibinfo {volume} {23}},\ \bibinfo {pages} {5048}
  (\bibinfo {year} {1981})}\BibitemShut {NoStop}%
\bibitem [{\citenamefont {Levinea}()}]{resistor}%
  \BibitemOpen
  \bibfield  {author} {\bibinfo {author} {\bibfnamefont {J.~D.}\ \bibnamefont
  {Levinea}},\ }\href@noop {} {\bibfield  {journal} {\bibinfo  {journal} {C R C
  Critical Reviews in Solid State Sciences}\ }\textbf {\bibinfo {volume}
  {5}}}\BibitemShut {NoStop}%
\bibitem [{\citenamefont {Kaiser}\ \emph {et~al.}(2004)\citenamefont {Kaiser},
  \citenamefont {Chapman}, \citenamefont {Schlecht},\ and\ \citenamefont
  {Burghard}}]{CP723}%
  \BibitemOpen
  \bibfield  {author} {\bibinfo {author} {\bibfnamefont {A.~B.}\ \bibnamefont
  {Kaiser}}, \bibinfo {author} {\bibfnamefont {B.}~\bibnamefont {Chapman}},
  \bibinfo {author} {\bibfnamefont {U.}~\bibnamefont {Schlecht}}, \ and\
  \bibinfo {author} {\bibfnamefont {M.}~\bibnamefont {Burghard}},\ }\href@noop
  {} {\bibfield  {journal} {\bibinfo  {journal} {AIP Conf. Proc.}\ }\textbf
  {\bibinfo {volume} {723}},\ \bibinfo {pages} {99} (\bibinfo {year}
  {2004})}\BibitemShut {NoStop}%
\bibitem [{\citenamefont {Brankovic}\ \emph {et~al.}(2007)\citenamefont
  {Brankovic}, \citenamefont {Brankovic}, \citenamefont {Bernik},\ and\
  \citenamefont {Zunic}}]{ZnO}%
  \BibitemOpen
  \bibfield  {author} {\bibinfo {author} {\bibfnamefont {Z.}~\bibnamefont
  {Brankovic}}, \bibinfo {author} {\bibfnamefont {G.}~\bibnamefont
  {Brankovic}}, \bibinfo {author} {\bibfnamefont {S.}~\bibnamefont {Bernik}}, \
  and\ \bibinfo {author} {\bibfnamefont {M.}~\bibnamefont {Zunic}},\
  }\href@noop {} {\bibfield  {journal} {\bibinfo  {journal} {Journal of the
  European Ceramic Society}\ }\textbf {\bibinfo {volume} {27}},\ \bibinfo
  {pages} {1101} (\bibinfo {year} {2007})}\BibitemShut {NoStop}%
\end{thebibliography}%

\end{document}